\documentclass[
    reprint,
    aps,
    amsmath,
    amssymb,
    superscriptaddress]{revtex4-2}
\usepackage{graphicx} 
\usepackage{dcolumn}  
\usepackage{bm}       
\usepackage{amsfonts}
\usepackage{dsfont}
\usepackage{amsmath}
\usepackage{url}
\begin{document}

\title{Three-temperature modeling of laser-induced damage process in silicon}

\author{Prachi Venkat}
\affiliation{%
Kansai Photon Science Institute, National Institutes for Quantum  Science and Technology (QST), Kyoto 619-0215, Japan }

\author{Tomohito Otobe}
\email[]{otobe.tomohito@qst.go.jp}
\affiliation{%
Kansai Photon Science Institute, National Institutes for Quantum  Science and Technology (QST), Kyoto 619-0215, Japan }
\begin{abstract}
Laser excitation in silicon from femto- to pico-second time scales is studied. We assume the Three-Temperature Model (3TM)   which describes the dynamics of the distinct quasi-temperatures for electrons, holes, and lattice. Numerical results for damage threshold reproduce the experimental results not only quantitatively, but qualitatively as well,  showing dependence on laser pulse duration. Comparison with experimental data suggests that electron emission and thermal melting are both responsible for damage in silicon. We found that electron-phonon  relaxation time has a significant effect on pulse duration dependence of electron emission.
\end{abstract}
\maketitle

In the recent years, we have been able to use intense laser pulses of duration ranging from femto- to pico-second time scales to investigate laser-matter interaction. In particular, laser processing of semiconductors using ultrashort pulses attracts great interest, due to its application in nano-structuring \citep{Gattass2008,Stoian-2020}. 
Ultrafast laser pulse enables precise processing of the material, without causing damage in the surrounding area \citep{Sugioka2014}.

Laser processing involves various aspects of physics such as optics, quantum mechanics, material science and thermodynamics, to name a few. Thus, numerical modeling is crucial in order to understand and optimize the process.  
Numerical modeling of laser excitation has seen considerable interest over the years \citep{Agassi-1984,VanDriel-1987,Chen-2005,Popov-2011,Lipp-2014,Tsibidis-2012}. In particular, the processing of silicon has important application, not only in understanding the fundamental science but also industrially \cite{Sokol-1995,Sokol-1998}.

When the laser pulse is shorter than the time-scale of phonon excitation, electronic excitation by photo-absorption is the dominant  process. The non-equilibrium phase with hot electron-hole pairs and cool lattice relaxes into thermal equilibrium via electron-phonon interaction.

In order to study non-equilibrium dynamics of electrons and lattice, we have to develop a numerical model including dynamics of electro-magnetic fields, electrons, lattice, and the interaction among them.
The Two-Temperature Model (TTM) has been proposed by Anisimov \textit{et al.} to describe the electron-lattice coupling in metals \citep{Anisimov1974}. 
The TTM has been extended to study excitation in semiconductors \cite{VanDriel-1987} and different models have been developed, one of them being the self-consistent  density-dependent TTM  (nTTM) \cite{Chen-2005}.
One of the critical approximations in TTM for semiconductors is assuming thermal equilibrium in electrons.
There are various differences in the TTMs for semiconductors found in existing literature, such as the treatment of excitation process and the values of optical parameters of silicon \cite{Agassi-1984, Lipp-2014,Tsibidis-2012,Ramer-2014}.

It has been observed that electron and hole relaxation in conduction and valence bands is faster than relaxation of the system as a whole \citep{Zurch2017} . As the electron temperature increases, electron-hole scattering frequency decreases significantly \cite{Terashige-2015}. The calculations presented using first-principles numerical simulation show that electrons and holes have different energy, depending on the band structure. The energy difference between electrons and holes suggests that quasi-temperatures are different in the conduction band (CB) and valence band (VB) \citep{Otobe-2019}. Also, the decreased electron-hole scattering frequency indicates that it is important to take quasi-temperatures of electrons and holes into consideration.

In this work, we would like to propose a modified  nTTM treating the evolution of  quasi-temperatures in CB, VB and lattice (3TM). 
A similar study has been presented by Silaeva \textit{et al.} \citep{Silaeva_2012}, in order to study laser assisted atom probe tomography.

The 3TM is modeled to consider three sub-systems, electrons, holes and lattice, and their temperatures are calculated separately. The carrier densities are also calculated distinctively for electrons and holes. We treat the dynamics of laser field employing the one-dimensional Maxwell's equations, which are solved using the Finite Difference Time Domain (FDTD) approach.
In order to understand and define damage in silicon, we assume several processes that may be responsible for structural changes and compare their calculated thresholds with the experimental damage thresholds.


The 3TM is similar to the previous nTTM formalisms \citep{Chen-2005} in its formulation. The major difference lies in the separate evolution of electron and hole temperatures and densities. 3TM also considers the changes in optical parameters of silicon during the excitation processes which in turn, affect the evolution of carrier densities and temperatures.
The single and double photon excitation processes have been modified from previous studies to include the effect of band re-normalization. Drude model is used to calculate dielectric function, including the effect of electron-hole-phonon collisions.

The time-evolution of electron and hole densities, $n_e$ and $n_h$ is described as:
\begin{equation}
    \begin{split}
    \frac{\partial n_{e(h)}}{\partial t} & = \frac{\alpha I}{\hbar \omega_0} + \frac{\beta I^2}{2\hbar \omega_0} - \gamma_e n_en_en_h -\gamma_h n_hn_hn_e \\
     & + \frac{1}{2}(\theta_e n_e + \theta_h n_h) + \nabla D_{e(h)}\cdot\vec J_{e(h)} + D_{e(h)}\nabla\cdot \vec J_{e(h)} \\
     & -(+) \mu_{e(h)}\nabla\cdot n_{e(h)}\vec{F} -(+) \mu_{e(h)}\nabla n_{e(h)}\cdot\vec{F}\\ & -(+) \mu_{e(h)} n_{e(h)} \nabla\cdot \vec{F}
    \end{split}\label{source}
\end{equation}
where $\omega_0$ is the laser frequency and $\alpha$ is the single photon absorption coefficient for transition from VB to CB \citep{Green2008}. $\beta$ is the two-photon absorption coefficient for which we use the DFT calculation when $2\hbar\omega_0 > E_g$ \cite{Murayama-1995} and around the band gap energy we employ interpolation to the model described in Ref.\cite{Alan2007}. $\gamma_{e(h)}$ is the Auger re-combination coefficient \citep{Silaeva_2012} and $\theta_{e(h)}$ is the impact ionization coefficient \citep{Chen-2005}. Equation \ref{source} also includes the effect of spatial charge distribution and the associated electric field, and $J_{e(h)}, D_{e(h)}$ and $\vec{F}$ are the charge current, diffusion coefficient and the electric field induced by the electron--hole separation, respectively \cite{Suppmat}.

The total dielectric function along with the effect of band structure re-normalization \citep{Sokol-2000} is expressed by 
\begin{equation}
    \epsilon(\omega) = 1 + \frac{n_0 - n_e}{n_0}\epsilon_L(\omega + \delta E_g/\hbar) + \epsilon_D(\omega) 
\end{equation}
where $n_0$ is the density of valence electrons. It should be noted that $n_e$ and $n_h$ are nearly the same due to the effect of $\vec{F}$ and can be approximated at $n_e$. $\epsilon_L(\omega)$ is the innate dielectric function, $\delta E_g$ represents the band re-normalization by carrier density, and $\epsilon_D$ is the complex  dielectric function calculated from Drude model.
The temperature dependent optical parameters of silicon are referred to from Ref.\citep{Green2008}.
$\epsilon_D$  accounts for the effect of plasma in the excited system. 
Considering the electron and hole sub-systems,
\begin{equation}
    \begin{split}
        \epsilon_D(\omega) = & -\frac{4\pi n_e e^2}{\omega_0^2}\left[ \frac{1}{m^*_e\left(1 + i\frac{\nu_e}{\omega_0}\right)} + \frac{1}{m^*_h\left(1 + i\frac{\nu_h}{\omega_0}\right)} \right]
    \end{split}
\end{equation}
where $m^*_{e(h)}$ is the effective mass and  $\nu_e$ and $\nu_h$ are the collision frequencies, describing the electron-hole ($e-h$), electron-phonon ($e-ph$) and hole-phonon ($h-ph$) collisions. The $e-ph$ and $h-ph$ collisions are assumed to have the same frequency which is dependent on lattice temperature \citep{Ramer-2014}. Effect of electron-ion core collisions is also considered \cite{Sato-2014,Suppmat}. The collision frequencies for $e-h$ interactions are calculated as per the model presented in Ref.\citep{Terashige-2015}.
The total one-photon absorption coefficient including free-carrier absorption is 
\begin{equation}
  \alpha_f=\frac{2\pi}{\lambda}\Im[\sqrt{\epsilon(\omega)}],
   \end{equation} 
where $\lambda$ is the laser wavelength.

Since 3TM considers three sub-systems, \textit{viz.} electron, hole and lattice, their temperatures also evolve separately.
The temperature evolution is expressed as:
\begin{equation}
\begin{split}
    C_{e(h)}\frac{\partial T_{e(h)}}{\partial t}= & m_{r,{e(h)}}(\alpha_f I + \beta I^2) +E_g\gamma_{e(h)} n_{e(h)} n_{e(h)} n_{h(e)} \\ & -\frac{C_{e(h)}}{\tau}(T_{e(h)}-T_l)
-\nabla \cdot \vec{w}_{e(h)} \\ & -\frac{\partial n_{e(h)}}{\partial t} \left(m_{r,{e(h)}}E_g 
+ \frac{3}{2}k_B T_{e(h)} H_{-1/2}^{1/2}(\eta_{e(h)}) \right) \\ & - m_{r,{e(h)}}n_{e(h)}\left( \frac{\partial E_g}{\partial T_l}\frac{\partial T_l}{\partial t}
+  \frac{\partial E_g}{\partial n_{e(h)}}\frac{\partial n_{e(h)}}{\partial t}\right),
\end{split}
\label{Te}
\end{equation}

\begin{equation}
    C_l\frac{\partial T_l}{\partial t}=-\nabla\cdot(\kappa_l \nabla T_l)+\frac{C_e}{\tau}(T_e-T_l)+\frac{C_h}{\tau}(T_h-T_l).
    \label{Tl}
\end{equation}
The third and fourth terms in Eq.~(\ref{Te}) account for the loss of energy due to electron-lattice interaction and energy current. The last two terms on right hand side include the changes in carrier density and band gap energy. Here, $H_{\xi}^{\zeta}(\eta)=F_{\zeta}(\eta)/F_{\xi}(\eta)$ and  $F_{\xi}(\eta)$ is the Fermi integral \cite{Suppmat}. The heat capacities $C_{e(h)}$ are calculated from the carrier densities and temperatures \cite{Suppmat}. 
$T_l$ is calculated following the empirical model, where the term for carrier temperature is replaced by terms for electron and hole temperatures, as described in Eq.~ (\ref{Tl}). Here, $\kappa_l$ is the thermal conductivity \cite{Suppmat}.

The propagation of the laser pulse is described by solving the Maxwell's equations using FDTD method. Mur's absorbing boundary condition is employed to prevent reflection from the boundary \citep{Mur-1981}. Another salient feature is that the electric field is considered to be complex for the calculation of laser intensity, to ensure a non-zero field at all points in time and space. Assuming a one-dimensional system, the electric field is:
\begin{equation}
    E(x,t) = E_0(x,t)\exp[i\omega t]
\end{equation}
Here $E_0$ includes Gaussian envelope defining the shape of the pulse as $\exp[-(t-4T)^2/T^2]$, where $T=t_p/(4\sqrt{\ln 2})$, $t_p$ being the FWHM pulse duration.
\begin{equation}
    I(x,t) = \frac{c}{8\pi}\Re[\sqrt{\epsilon}]|E_{0}(x,t)|^2
\end{equation}
where $I(x,t)$ is the laser intensity. 
Evaluation of charge current induced by the laser field is a crucial part of the module. We calculate the current with and without excitation i.e., for photo-absorption and dielectric response. For dielectric response, $j_0(x,t)$ is calculated as:
\begin{equation}
    j_0(x,t) = \chi_r(\omega)\frac{\partial P(x,t)}{\partial t} = -\chi_r(\omega)\frac{\partial^2A(x,t)}{c\partial t^2}
\end{equation}
where $A(x,t)$ is the vector potential,  $P$ is the polarization and $\chi_r$ is the real part of susceptibility ($\chi = (\epsilon-1)/4\pi$).
The Maxwell's equation thus, becomes 
\begin{equation}
   \frac{1}{c^2}(1+4\pi\chi_r(\omega))\frac{\partial^2A(x,t)}{\partial t^2} - \frac{\partial^2A(x,t)}{\partial x^2} = \frac{4\pi}{c}j(x,t)
\end{equation}
where
\begin{equation}
    j(x,t) = (\alpha_f(\omega) + \beta(\omega)I(x,t))\frac{c\Re[\sqrt{\epsilon}]}{4\pi}E(x,t)
\end{equation}
is the current associated with photo-absorption.

The time evolution of 3TM is calculated using $4^{th}$ order Runge-Kutta method. Euler's method is used in the FDTD method to solve Maxwell's equations. In our study, the 3TM time step ($dt$) is of the order $10^{-1}$~fs.  For FDTD calculation, we define the another time step , $dt_m\sim 10^{-3}$~fs, because the time-scale of electromagnetic field dynamics is faster than that of electron dynamics. For spatial parameters, the simulation grid size is taken to be 150~\AA ~ and the silicon film thickness is 60.5 $\mu$m, which is thicker than the penetration depth, which is $\sim$10 $\mu$m for 775~nm laser.

\begin{figure}[h!]
    \includegraphics[width=0.5\textwidth]{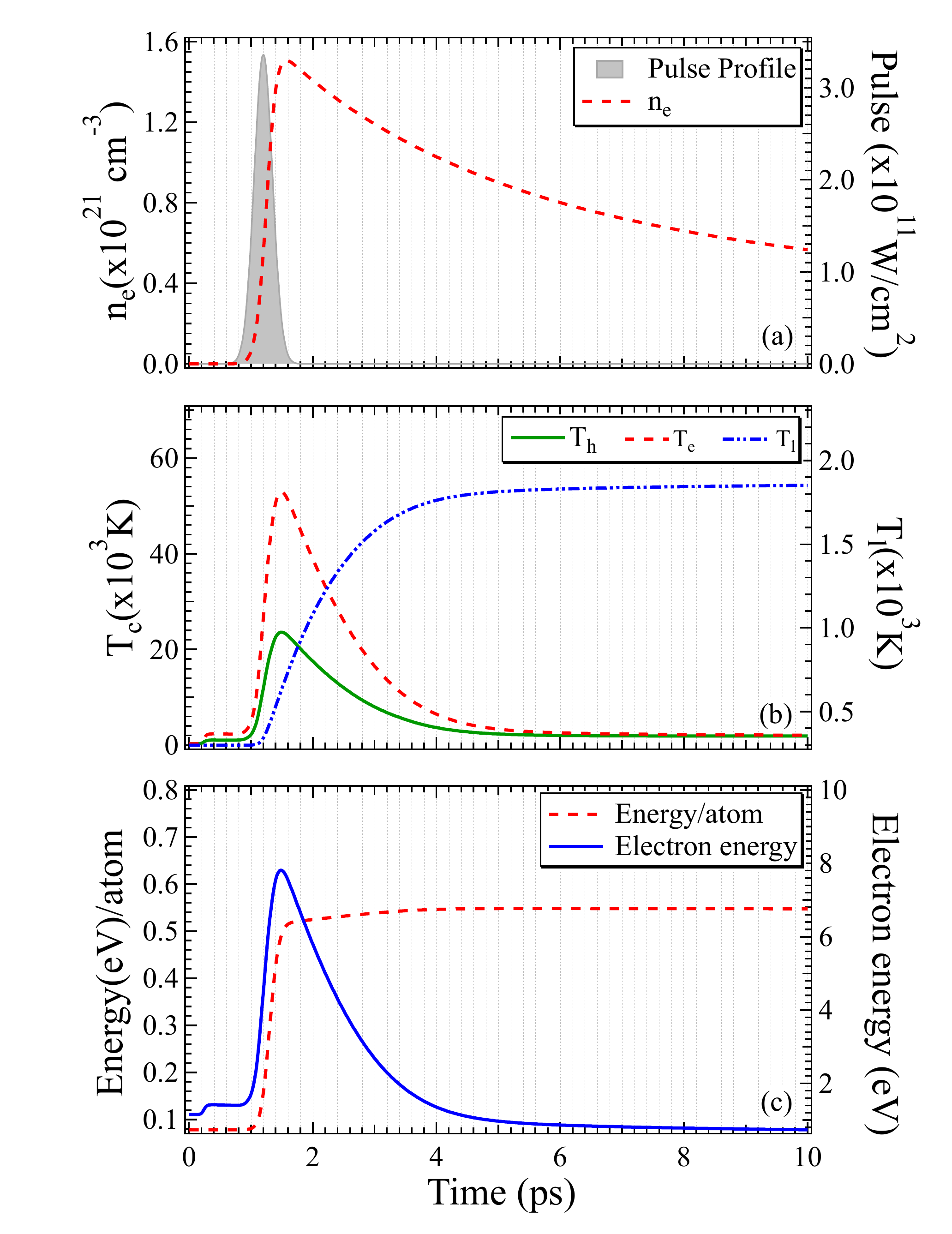}
    \caption{Time evolution of (a) electron density, (b) $T_e$, $T_h$, $T_l$, (c) total energy per atom and electron energy.}
    \label{timevol}
\end{figure}

Figure \ref{timevol} shows the time evolution of electrons, holes and the lattice dynamics in silicon surface for interaction with a 775~nm, 350~fs pulse having incident fluence 0.18~J/cm$^2$. We assume the one-dimensional system for 3TM.
Figure \ref{timevol}(a) shows the electron density, along with the pulse profile in the background. The steep rise in $n_e$ is due to photo-absorption, followed by fall in  density due to Auger re-combination.  The electron, hole and lattice temperature evolution in Fig.~\ref{timevol}(b) shows that the three sub-systems eventually reach equilibrium at $\sim 1.85\times10^3$~K. Initially the carrier temperatures rise steeply due to photo-absorption and subsequent plasma heating. This is followed by re-combination and carrier-phonon dynamics which re-distribute the energy. There is transfer of energy from electrons and holes to the lattice, causing the lattice temperature to rise. Figure \ref{timevol}(c) shows the evolution of electron energy and the total energy per atom. Since the penetration depth for the pulse wavelength of  775~nm pulse is high,
the gradients of $n_{e(h)}$ and $T_{e(h)}$ are small. Small $\nabla n_e(r)$ and $\nabla T_e(r)$ suggest low charge and energy current. Therefore, the total energy is nearly constant with time, which also demonstrates the stability of the 3TM simulation. 

The damage in a semiconductor is defined in different ways throughout the published literature \cite{Allen-2003,Bonse-2002,Izawa_2006,Mirza2016,Thorstensen-2012}.
Broadly speaking, the structural and phase changes observed in the lattice of laser excited material are defined as 'damage'. More specifically, there are different processes leading to damage, depending on the conditions of laser excitation and material properties.

One of the most prominent processes for damage is thermal melting.
The carrier-phonon interactions lead to re-distribution of energy among the carriers and lattice, causing the lattice temperature to rise up to melting point,  which can manifest as damage in the material \citep{Terashige-2015}.
Thermal melting threshold can be studied from two perspectives, initial commencement of melting which leads to what is known as the 'slush' phase (partial melting) \citep{Agassi-1984}, and the completion of melting in crystalline silicon (c-Si) when bond configuration changes to tetrahedral after absorption of latent heat from the carriers (complete melting). 
The two different approaches to study melting in silicon \citep{Korfiatis-2007} lead to different understanding of the onset of damage. 

Critical density has also been considered as a crucial parameter to study damage \cite{jurgens-2019} and the threshold is considered when the electron density reaches the critical density and plasma state exists in the material. The Drude-model indicates an abrupt increase in absorbed energy above the critical density threshold, which leads to sudden increase in melting thresholds.

The first-principles calculation by the time-dependent density functional theory (TDDF) indicates that the ablation threshold of $\alpha$-Quartz lies between melting and cohesive energy \cite{Sato-2015}. The TDDFT results suggests the bond-breaking energy is a reasonable candidate for the damage threshold.
 Threshold for breaking of bonds is  calculated as the threshold fluence when energy per atom reaches 2.3~eV, which is half the cohesive energy of a silicon crystal \cite{Lutrus-1993}. 
 
Emission of electrons from the surface into vacuum is also an important candidate for understanding damage. When the electron temperature is high, electron emission (e-emission) from the surface induces an impulsive force, Coulomb explosion being the extreme case \cite{Roeterdink-2003}. We estimate the e-emission threshold   as the threshold when the electron energy exceeds the work function for silicon (4.65~eV).

\begin{figure}[h!]
    \includegraphics[width=0.5\textwidth]{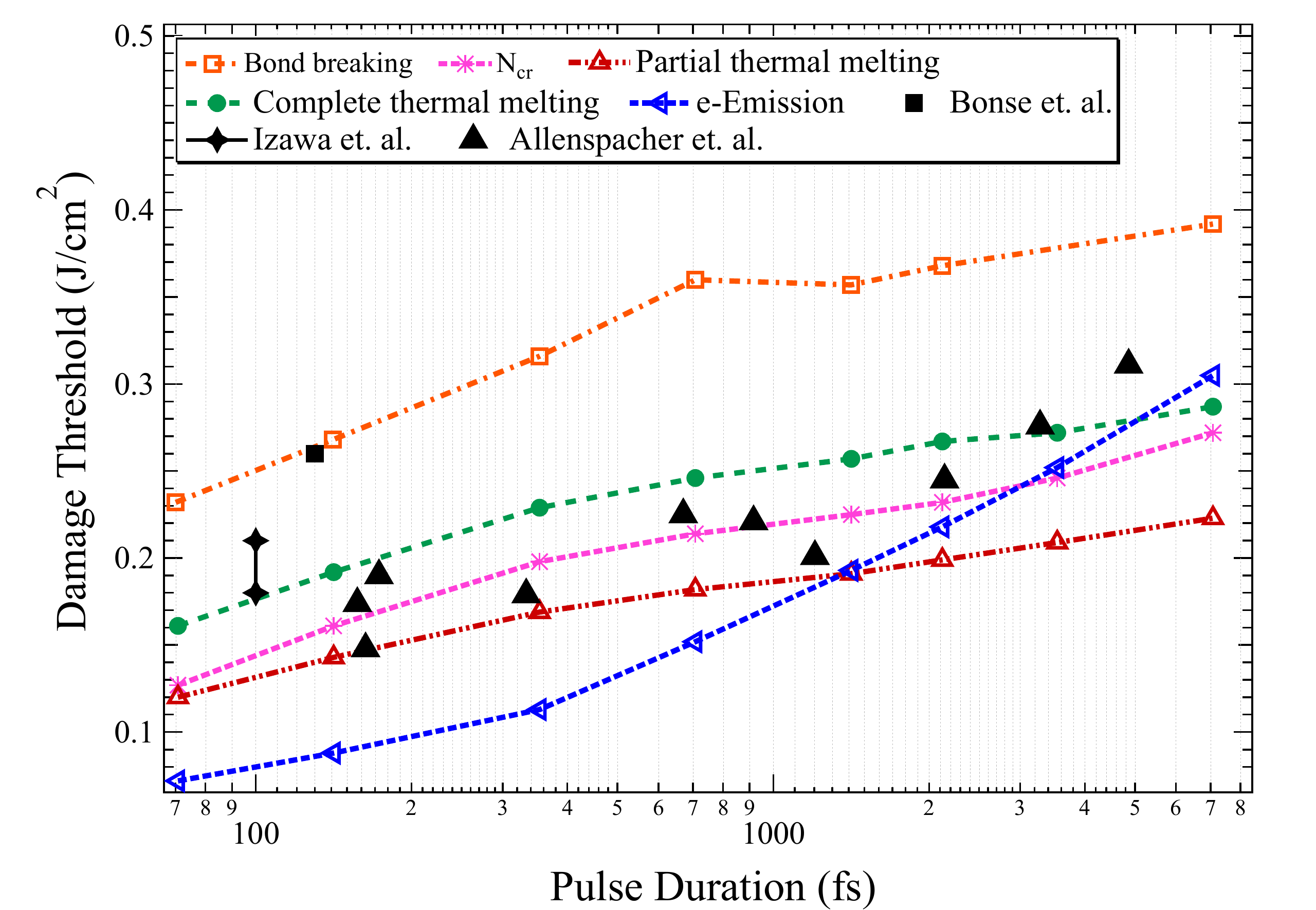}
    \caption{Comparison of damage thresholds for different processes with the experimental threshold data \citep{Allen-2003,Bonse-2002,Izawa_2006}.}
    \label{Allen775}
\end{figure}

Figure~\ref{Allen775} shows the  numerical results for damage threshold by 3TM with a 775~nm laser. Experimental data is also included in the figure. Data from Allenspacher \textit{et al.} shows results for damage threshold in laser processed silicon using 775~nm pulse of varying duration \citep{Allen-2003}, Bonse \textit{et al.} data shows ablation damage threshold for a 800~nm pulse of duration 130~fs \citep{Bonse-2002} and Izawa \textit{et al.} data shows the crystallization threshold range for a $800$ nm pulse of duration 100~fs \citep{Izawa_2006}. 
The damage in these experimental studies are defined differently, which is why there is a considerable difference in the threshold data. Allenspacher \textit{et. al.} \cite{Allen-2003} observed morphology of the surface and  threshold was estimated as the  damage probability becomes zero. 
Bonse \textit{et. al.} \cite{Bonse-2002} defined damage threshold as the fluence at which area of damaged region becomes zero by extrapolating the fitted line. Izawa \textit{et al.} \cite{Izawa_2006} defined a range of incident fluence for crystallization of amorphous silicon (a-Si) on c-Si. 

In Fig.~\ref{Allen775},  the experimental data suggests weak $t_p$-dependence in the short pulse regime (SPR, $t_p<1$~ps), while it is sensitive to $t_p$ in the long pulse regime (LPR, $t_p>1$~ps).
In SPR, experimental results for threshold lie between partial melting and complete melting. At $\sim1$~ps, there is a reversal in the partial melting and e-emission threshold, and e-emission coincides with the experimental results in LPR.
The $t_p$-dependence for LPR indicates a power law of $t_p^\eta$, which has also been reported with fused silica and CaF2 ($\eta=1/2$) \cite{stuart-1995}.
In the LPR, the experimental data fits well with a $t_p^{0.15}$ dependence, and the e-emission threshold fits with a similar power law of $t_p^{0.14}$.
The agreement of e-emission threshold with experimental data indicates that the impulsive Coulomb force induces damage above 1~ps.
 The $t_p$ dependence of threshold suggests that thermal melting and electron emission are necessary and sufficient conditions for causing damage in silicon. 

A possible explanation to the damage process is a cooperative relationship between bond softening by  excitation and Coulomb force due to electron emission.
The melting threshold indicates Si-Si bond-softening, while electron emission indicates the presence of impulsive Coulomb force. Although the melting process takes time, the impulsive Coulomb force can enhance the lattice dynamics. On the other hand, Coulomb force may not be strong enough to break the cold Si-Si bonds. 
The thermal melting and e-emission thresholds cross at $\sim1.4$~ps, indicating that both these processes are together responsible for causing damage in silicon.


The amorphization threshold of Bonse \textit{et al.} coincides with the calculated threshold for breaking of bonds. The Izawa \textit{et al.} results give a range of crystallization for amorphous silicon. As it can be seen in Fig.~\ref{Allen775}, the threshold for crystallization lies between the melting and bond-breaking threshold.


\begin{figure}[h!]
    \includegraphics[width=0.5\textwidth]{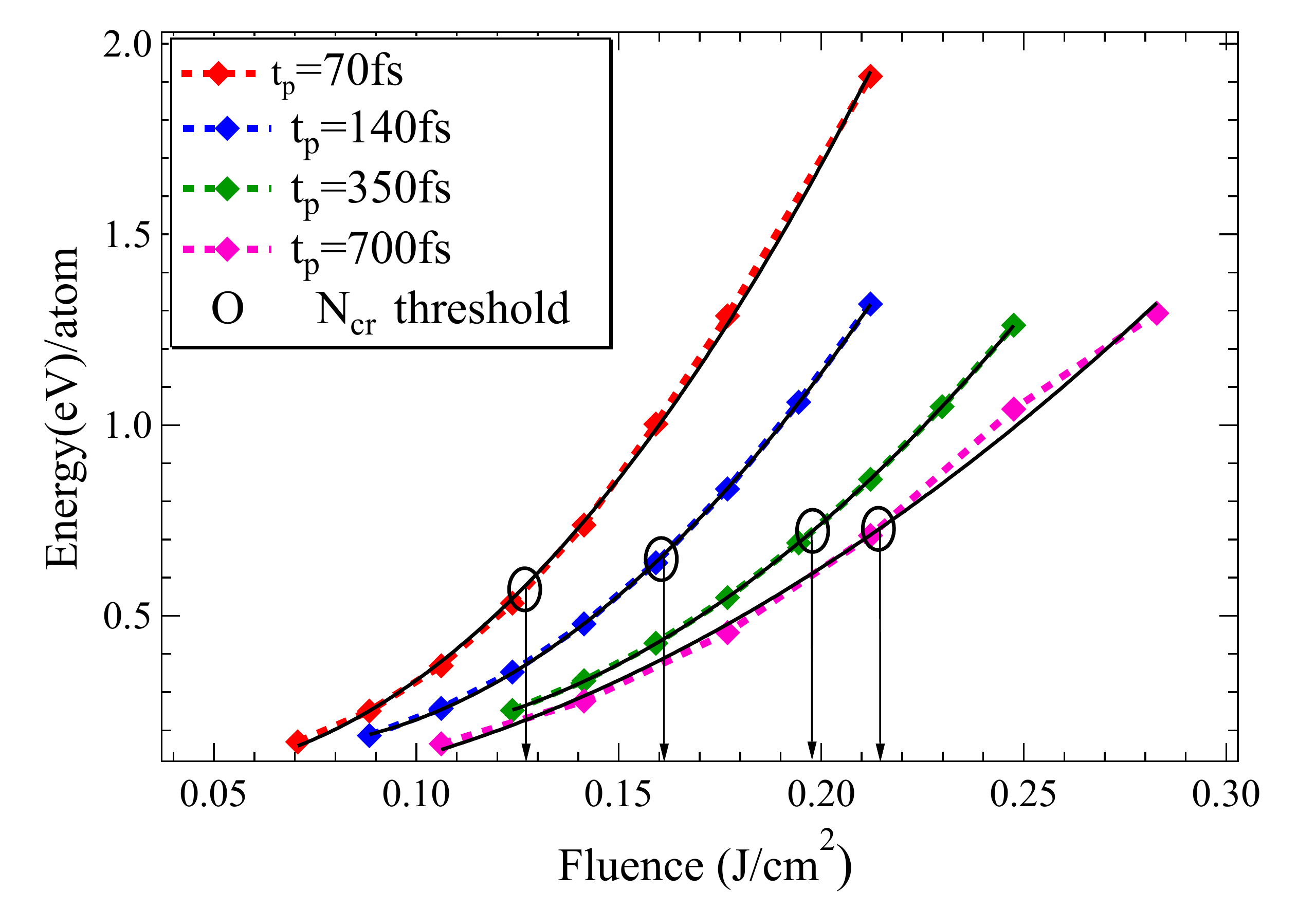}
    \caption{ Energy per atom at different incident fluences, plotted for different pulse durations.Thresholds for critical density ($1.86\times10^{21}$~cm$^{-3}$) for different pulse durations are also indicated. }
    \label{ev_atom_comp}
\end{figure}

Although considered as an important measure of studying damage, we find that in our model critical density does not qualify as the dominant reason for damage in silicon. 
When the plasma frequency increases beyond laser frequency, photo-absorption by free carriers is assumed to be significant.
As it can be seen in Fig.~\ref{ev_atom_comp}, the laser fluence dependence of the absorbed energy shows monotonic increase for various pulse duration. However, we found that the fluence dependence can be fitted by the quadratic function (black solid lines in Fig.~\ref{ev_atom_comp}). The quadratic increase of the energy above the critical density thresholds indicates that the critical density cannot be a favorable criterion for laser damage.


\begin{figure}[h]
    \includegraphics[width=0.5\textwidth]{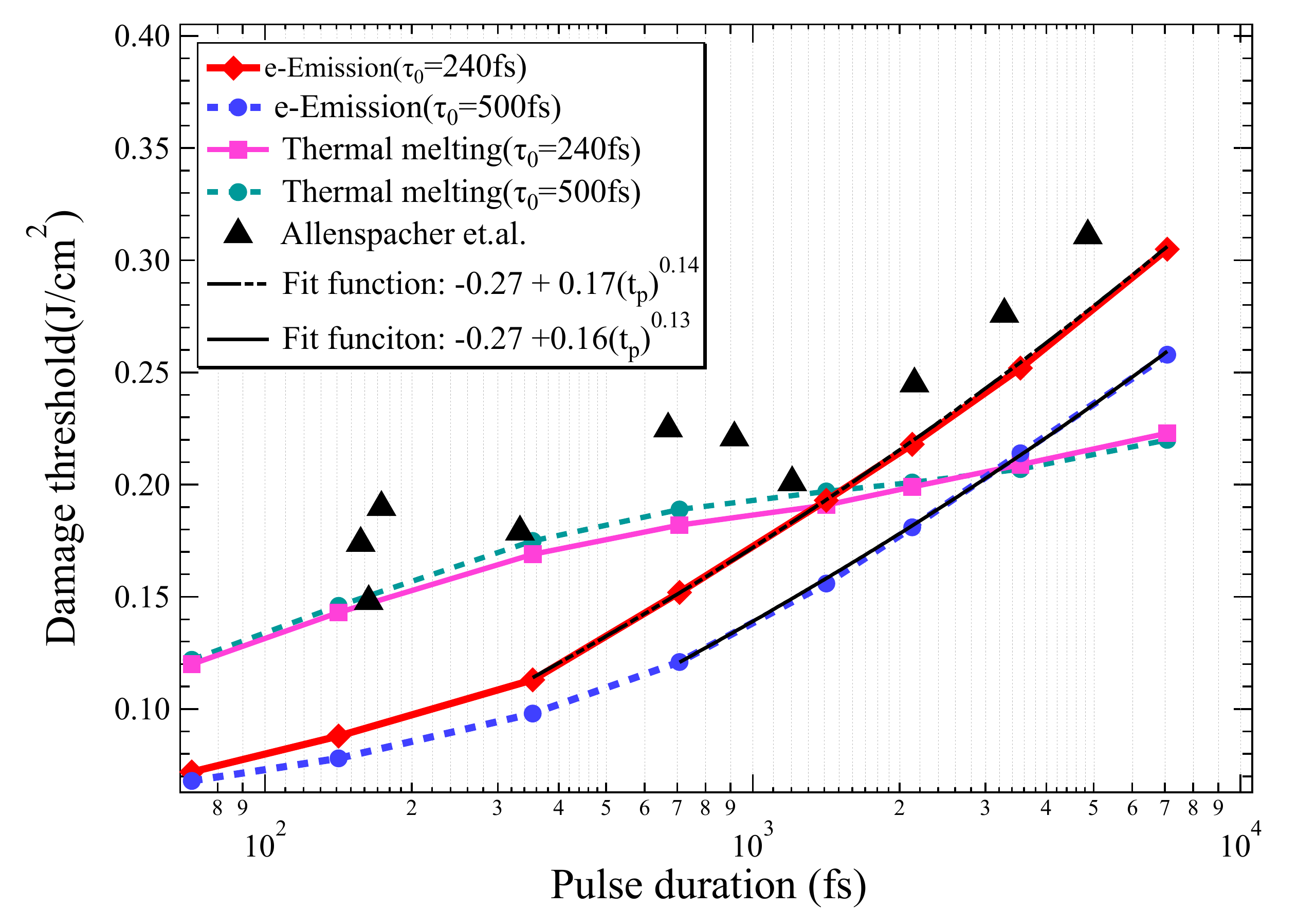}
    \caption{Comparison of electron emission and thermal melting thresholds for $\tau_0=240$ fs and $500$ fs.}
    \label{tau_comp}
\end{figure}

The e-emission is an important process in LPR and its corresponding threshold indicates a significant pulse duration dependence, as compared to thresholds related to melting. Since $T_e$ defines the e-emission threshold, it should depend on the $e-ph$ relaxation time.
Figure \ref{tau_comp} shows the threshold for thermal melting and e-emission for different $e-ph$ relaxation time $\tau=\tau_0[1 + \{n_e/8\times10^{20}( 1/$cm$^3)\}^2]$, by taking the constant $\tau_0$ as $\tau_0=240$~fs and 500~fs. The e-emission threshold shows a qualitative shift towards higher $t_p$ dependence, which also depends on $\tau_0$. The threshold can be studied in two parts, where higher dependence is seen for longer $t_p$. 
We present the fitted line with $t_p^{\eta}$ for e-emission threshold in LPR, where $\eta=0.14$ and $0.13$ for $\tau_0=240$~fs and $500$~fs, respectively. The deviation from the fitted line occurs at $1.5\tau_0$ for $\tau_0=240$~fs and $1.4 \tau_0$ for $\tau_0=500$~fs, respectively.
The crossing points of thermal melting and e-emission threshold occur at $5\tau_0$ and $6\tau_0$ for $\tau_0=240$~fs and 500~fs. Due to slower energy transition in case of $\tau_0=500$~fs, the crossing point shifts higher. 
Also, since the $t_p$ dependence follows almost the same power law for both cases of $\tau_0$, it shows that the physical process is almost the same when $t_p$ is sufficiently long. 
The threshold for thermal melting is not affected by the change in $\tau_0$, suggesting that the temperature dependence of photo-absorption by electron-hole system does not affect the lattice temperature significantly. 

In this work, we studied the laser excitation and possible damage processes in silicon numerically using 3TM combined with Maxwell's equations. The model allowed us to study in detail the excitation dynamics and energy re-distribution in silicon for electrons, holes and the lattice. Pulse duration dependence of the damage threshold was also studied in detail. The 3TM simulations reproduced the experimental damage thresholds and the qualitative dependence on incident pulse duration. The comparison of experimental threshold alongside the calculated thresholds indicated that thermal melting and e-emission are both important processes, contributing to damage in silicon. However, critical density threshold is not an equitable criterion for damage. Also, variation in the $e-ph$ relaxation time showed that  $e-ph$ interaction dynamics plays an important role in damage process with longer pulse duration. 

\acknowledgments
\noindent  This research is supported by MEXT Quantum Leap Flagship Program (MEXT Q-LEAP)
under Grant No.  JPMXS0118067246. 
This research is also partially supported by JST-CREST under Grant No. JP-MJCR16N5.
The numerical
calculations are carried out using the computer facilities of the
SGI8600 at Japan Atomic Energy Agency (JAEA).





%

\end{document}